\documentclass[conference]{IEEEtran}
\IEEEoverridecommandlockouts
\usepackage{cite}

\usepackage{cite}
\usepackage{amsmath,amssymb,amsfonts}
\usepackage{amsthm}
\usepackage{algorithmic}
\usepackage{graphicx}
\graphicspath{{pdf/}}
\DeclareGraphicsExtensions{.pdf,.jpeg,.png}
\usepackage{textcomp}
\usepackage{xcolor}
\usepackage{algorithm}
\usepackage{siunitx}
\usepackage{url}
\usepackage{tikz}
\usepackage{mdframed}

\newtheorem{alg}{Algorithm}

\newtheorem{remark}{Remark}

\usepackage{xspace}
\usepackage{bm}

\newcommand{\R}{\ensuremath{\mathbf{R}}\xspace}

\newcommand{\est}[1]{\ensuremath{\hat{#1}}}

\newcommand{\mac}[1]{\ensuremath{\boldsymbol{#1}}\xspace}
\newcommand{\kol}[2][]{\ensuremath{\boldsymbol{#2}_{#1}}\xspace}
\newcommand{\wek}[1]{\ensuremath{\boldsymbol{#1}}\xspace}
\newcommand{\ele}[2][]{\ensuremath{#2_{#1}}}
\newcommand{\ev}[1][]{\wek{e}}

\newcommand\copyrighttext{%
  \footnotesize \textcopyright 2023 IEEE.\@ Personal use of this material is permitted.
  Permission from IEEE must be obtained for all other uses, in any current or future
  media, including reprinting/republishing this material for advertising or promotional
  purposes, creating new collective works, for resale or redistribution to servers or
  lists, or reuse of any copyrighted component of this work in other works.%
}
\newcommand\copyrightnotice{%
\begin{tikzpicture}[remember picture,overlay]
\node[anchor=south,yshift=10pt] at (current page.south) {\fbox{\parbox{\dimexpr\textwidth-\fboxsep-\fboxrule\relax}{\copyrighttext}}};
\end{tikzpicture}%
}
\begin{document}
%

\title{PCA-aided calibration of systems comprising multiple unbiased sensors%
\thanks{This work was supported by the National Centre for Research and Development of Poland - project No. DOB-SZAFIR/03/B/018/01/2021.}%
}

\author{\IEEEauthorblockN{Marek W. Rupniewski}
    \IEEEauthorblockA{%
        \textit{Institute of Electronic Systems}\\
        \textit{Warsaw University of Technology}\\
        Nowowiejska 15/19, 00-665 Warsaw, Poland\\
        Email: Marek.Rupniewski@pw.edu.pl\\%
        ORCID: 0000-0003-3861-510X%
}}%

\maketitle
\copyrightnotice

\begin{mdframed}[linewidth=1.5pt]
    During my further study on the subject, I discovered that the algorithms
    presented in this paper are merely rediscoveries of those previously
    proposed by 
    Richard M. Voyles, J. Daniel Morrow, and Pradeep K. Khosla in
    their papers:
    \par
    \begin{enumerate}
        \item R. M. Voyles, J. D. Morrow, and P. K. Khosla ``Shape from motion
            approach to rapid and precise force/torque sensor calibration,'' in
            \emph{Proc. Int. Mech. Eng. Congr. Exp.}, Nov. 1995, pp. 2171–2175.
        \item R. M. Voyles, J. D. Morrow, and P. K. Khosla, ``Including sensor bias in
shape from motion calibration and sensor fusion,'' in \emph{Proc. IEEE Multi-
sens. Fus. Integrat. Conf.}, Dec. 1996, pp. 93–99.
    \end{enumerate}
\end{mdframed}
\begin{abstract}
    The calibration of sensors comprising inertial measurement units is crucial for reliable and accurate navigation. Such calibration is usually performed with specialized expensive rotary tables or requires sophisticated signal processing based on iterative minimization of nonlinear functions, which is prone to get stuck at local minima. We propose a novel calibration algorithm based on principal component analysis. The algorithm results in a closed-form formula for the sensor sensitivity axes and scale factors. We illustrate the proposed algorithm with simulation experiments, in which we assess the calibration accuracy in the case of calibration of a system consisting of 12 single-axis gyroscopes.

\end{abstract}


\section{Introduction}\label{sec:intro}
Accelerometers and gyroscopes are commonly known as inertial
sensors. Inertial measurement units typically consist of multiple such
sensors and are often augmented by magnetometers to estimate the inclination better. Before the navigation system is used, it must be calibrated.
It is especially vital for sensors produced in
micro-electrical-mechanical systems (MEMS) technology. Such sensors are
generally delivered uncalibrated~\cite{Sipos12} to reduce the production costs
of IMUs for mass-market products. There are two predominant classes of methods
for IMU sensor calibration. One is based on expensive specialized equipment,
such as precise mechanical platforms \cite{Titterton04} or optical tracking
systems (e.g., \cite{Wei22}). The other class, called multi-position calibration, relies on the measurements carried out under
static conditions, which utilizes the knowledge of the magnitude of the
measured vector quantity, such as the Earth's gravity, e.g.,~\cite{Nilsson14,Rohac15}. To the best of
our knowledge, the methods that fall into the latter class require nonlinear
optimization realized by iterative algorithms, e.g., Gauss-Newton algorithm \cite{Skog06}, Newton-Raphson \cite{Nilsson14}, Levenberg-Marquadt algorithm \cite{Sipos12}, or other algorithms provided by numerical toolboxes~\cite{Rohac15}.
Our study falls into multi-position calibration as well. However, we propose
measurement signal processing that leads to a closed form for the calibration
parameters. 
We consider systems that consist of $m$
single-axis sensors, each of which measures the projection of a given vector-valued quantity onto the sensor sensitive axis. Such systems may consist of, e.g.,
accelerometers, gyroscopes, or magnetometers. We assume the following
sensor measurement model
\begin{equation}
    \label{eq:readingmodel}
    \ele[ij]{m} = \kol[i]{a}^T \kol[j]{v} + \ele[ij]{\eta},\quad i=1,\,\dots,\,m,\quad  j=1,\,\dots,\,n,
\end{equation}
where $\ele[ij]{m}$ is the read-out of $i$-th sensor,
$\kol[j]{v}\in\R^d$ is the measured vector quantity at $j$-th position of the system,
$\kol[i]{a}\in\R^d$ is a vector that encodes the scale and the sensitive axis of the sensor,
and $\ele[ij]{\eta}$ is the Gaussian noise with zero mean. In our study, we treat dimension $d$ as an arbitrary positive integer. 
In practice, two values of $d$ are predominant, i.e., $d=3$ for the case of $3$-dimensional state space, and $d=2$ for the state space that takes the form of a~plane.
The absence of bias
terms in~\eqref{eq:readingmodel} may be an inherent property of the system's sensors, It can also result from bias estimation during a pre-calibration procedure, see Section~\ref{sec:model1}.
We assume that the measured quantity stays constant in magnitude for all $n$ positions of the considered system, i.e.,
\begin{equation}\label{eq:vnorm}
    \|\kol[j]{v}\| = c,\quad j=1,\,\dots,\,n,
\end{equation}
where $\| \kol[j]{v}\|$ denotes the Euclidean norm of vector $\kol[j]{v}$, and
$c$ is an arbitrary scalar constant. Earth's gravity or magnetic fields measured at a given point on Earth exemplify such quantities. The Angular rate of a rotary table that rotates with a fixed angular speed is another such quantity. In the former case, we may consider the calibration of a system consisting of accelerometers or magnetometers, and in the latter case --- a system comprising gyroscopes.

The paper is organized as follows. The next section presents the main
contribution of our study, i.e., the algorithm for finding the sensitive
vectors of a sensor set. Section~\ref{sec:positions}
discusses the number of positions required for the proposed calibration
procedure. Section~\ref{sec:model1} presents the results of numerical
experiments, in which we have used the proposed algorithm to calibrate 
a system that consists of four triads of single-axis gyroscopes, see Fig.~\ref{fig:model1}. Section~\ref{sec:conclusion} concludes our paper.
\IEEEpubidadjcol

\section{Calibration procedure}\label{sec:calibration}
Throughout the paper, we denote matrices with small bold letters with no
subscript, e.g., \mac{a}, matrix columns with single-subscripted bold letters, e.g.,
\kol[i]{a}, and matrix entries with regular double-subscripted letters, e.g.
\ele[ij]{a}.
Equation~\eqref{eq:readingmodel} takes the following form in the matrix notation.
\begin{equation}
    \mac{m} = \mac{p} + \mac{\eta} = \mac{a}^T \mac{v} + \mac{\eta}.\label{eq:macn}
\end{equation}
\subsection{Noiseless case}
Let us first consider the noiseless case, i.e., the case where $\mac{\eta}=\mac{0}$ in Eq.~\eqref{eq:macn}. Assume that matrix \mac{p} is of the maximum possible rank, which is~$d$. Thus, so are the ranks of matrices \mac{a} and \mac{v}.
Let vectors $\kol[1]{e}$, ..., $\kol[d]{e}\in\R^m$ constitute a linear basis of~the subspace $V\subset\R^m$ that is spanned by the columns of matrix~\mac{p}, and
let matrix \mac{b} define the decomposition of matrix \mac{p} columns in this basis, i.e., 
\begin{equation}
    \label{eq:bdef}
    \mac{p} = \mac{a}^T \mac{v} = \mac{e} \mac{b}.
\end{equation}
There must exist vectors \kol[1]{f}, ..., $\kol[d]{f}\in\R^d$ such that
\begin{equation}
    \label{eq:kolfdef}
    \kol[k]{e} = \mac{a}^T \kol[k]{f},\quad k=1,\,\dots,\,d.
\end{equation}
By combining Eqs.~\eqref{eq:bdef} and~\eqref{eq:kolfdef} we get
\begin{equation}
    \label{eq:vcompute}
    \mac{v} = \mac{f} \mac{b}
\end{equation}
and, in particular,
\begin{equation}
    \label{eq:vnormf}
    c^2 = \|\kol[j]{v} \|^2 = 
    \kol[j]{b}^T \mac{f}^T \mac{f} \kol[j]{b} =
    \kol[j]{b}^T \mac{g} \kol[j]{b}
    ,\quad j=1,\,\dots,\,n.
\end{equation}
We may treat Eqs.~\eqref{eq:vnormf} as a set of $n$ scalar equations for the
entries of a symmetric matrix $\mac{g} = \mac{f}^T\mac{f}$. Once the equation set is
solved for these entries, one may compute matrix \mac{f} by eigendecomposition of \mac{g}:
\begin{equation}
    \label{eq:fqlq}
    \mac{g} = \mac{f}^T\mac{f} = \mac{q}^T \mac{\lambda} \mac{q},
\end{equation}
and thus
\begin{equation}
    \label{eq:ff2f}
    \mac{f} = \mac{\lambda}^{\frac{1}{2}} \mac{q},
\end{equation}
where $\mac{\lambda}$ is a diagonal matrix with non-negative entries, and $\mac{q}$ is an orthogonal matrix. By combining Eqs.~\eqref{eq:vcompute} and~\eqref{eq:ff2f}
\begin{equation}
    \label{eq:vcomputef}
    \mac{v} = \mac{\lambda}^{\frac{1}{2}} \mac{q} \mac{b}.
\end{equation}
Eventually, by substituting~\eqref{eq:vcomputef} into~\eqref{eq:bdef} and solving it for \mac{a}, we get
\begin{equation}
    \label{eq:computea}
    \mac{a} = \mac{\lambda}^{-\frac{1}{2}}\mac{q}\mac{e}^T.
\end{equation}
The following algorithm concludes this subsection.
\begin{alg}\label{alg:noiseless}
    \par {\bf{Inputs:}} Noiseless sensor readings in the form of matrix
    $\mac{p}\in\R^{m\times n}$ of rank~$d$, the magnitude $c$ of the measured
    vector quantity (see Eqs.~\eqref{eq:readingmodel}--\eqref{eq:macn})
    {\bf{Output:}} Matrices $\est{\mac{v}}\in\R^{d\times n}$ and $\est{\mac{a}}\in\R^{d\times m}$ such that $\mac{p}=\est{\mac{a}}^T \est{\mac{v}}$.
    \begin{enumerate}
        \item Choose an arbitrary linear basis \kol[1]{e}, ..., \kol[d]{e} for
            the subspace spanned by the columns of matrix \mac{p} and decompose
            these columns relative to the basis:
            \[
                \mac{p} = \mac{e}\mac{b}.
            \]
        \item Solve the following set of linear equations for the entries of symmetric matrix $\mac{g}\in\R^{d\times d}$:
            \[
                c^2 = \kol[j]{b}^T \mac{g} \kol[j]{b},\quad
             j=1,\,\dots,\,n.
            \]
        \item Compute the eigendecomposition of matrix $\mac{g}$: 
            \[
                \mac{g} = \mac{q}^T \mac{\lambda} \mac{q}.
            \]
        \item Compute $\est{\mac{v}} = \mac{\lambda}^{\frac{1}{2}} \mac{q} \mac{b}$
            and $\est{\mac{a}} = \mac{\lambda}^{-\frac{1}{2}}\mac{q}\mac{e}^T$.
    \end{enumerate}
\end{alg}
\begin{remark}\label{rem:ortho}
    If Eq.~\eqref{eq:vnormf} has a unique solution \mac{g}, then columns of 
    matrices \est{\mac{v}} and \est{\mac{a}} are determined uniquely up to an orthogonal
    transformation, i.e., they are equal to \mac{v} and \mac{a}, respectively,
    up to the multiplication from the left by an arbitrary orthogonal
    matrix~$\mac{r}\in\R^{d\times d}$.
    If Eq.~\eqref{eq:vnormf} fails to have a unique solution \mac{g}, then
    the algorithm cannot recover the original matrix $\mac{v}$ from the
    readings even up to an orthogonal transformation.
\end{remark}
\begin{remark}\label{rem:c}
    If constant $c$ is not known, then Algorithm~\ref{alg:noiseless} cannot
    determine the scale factors of sensors. However, by taking any value of
    $c$, e.g., $c=1$, we may at least reconstruct the sensors' sensitive axes
    up to an orthogonal transformation, and find the scale factors of the
    sensors up to a common factor, provided that Eq.~\eqref{eq:vnormf} has a
    unique solution.
\end{remark}

\begin{figure}[tpb]
    \centering
    \includegraphics[]{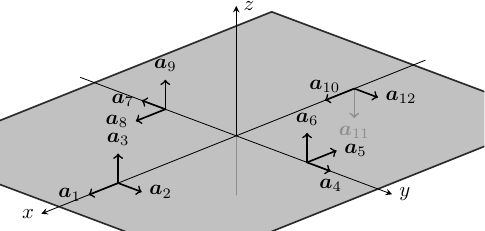}
    \caption{Sensitive vectors of four triads of gyroscopes considered in Section~\ref{sec:model1}}
    \label{fig:model1}
\end{figure}
\subsection{Noisy measurements}
Due to the noise terms, matrix \mac{m} of Eq.~\eqref{eq:macn} is of
the full rank, i.e., the probability of the rank of matrix \mac{m} being full is $1$.
Consequently, if the number of positions $n$ is bigger than the number of sensors $m$, then
the columns of matrix \mac{m} span the whole space~$\R^m$.
However, if the noise terms are small, the columns \kol[j]{m} treated as points in $\R^m$ must lie close to the subspace $V$ spanned by the columns of matrix \mac{p}.
Therefore, we may estimate the subspace $V$ as a $d$-dimensional subspace
$\est{V}\subset\R^m$ that is the closest to the columns of matrix \mac{m} in terms of the
mean squared Euclidean distance.
This task can be accomplished by Principal Component Analysis (PCA) method as
presented in Pearson's seminal paper~\cite{Pearson01}. If 
\begin{equation}
    \label{eq:truncpca}
    \mac{t} = \mac{m} \mac{w}, \quad \mac{w}\in\R^{m \times d}
\end{equation}
is the PCA transformation of $\mac{m}$ truncated to the first $d$ principal axes, then 
the columns of matrix \mac{t} span subspace $\est{V}$. Before we follow the
procedure introduced in the previous subsection, we need to approximate the
columns of matrix \mac{m} with those of matrix \mac{t}. Let us recall that the
truncated PCA transformation can be obtained by truncated Singular Value
Decomposition (SVD):
\begin{equation}
    \label{eq:truncsvd}
    \mac{m} \approx \mac{u}\mac{s}\mac{w}^T,\quad \mac{s}\in\R^{d\times d},
\end{equation}
where the columns of matrices $\mac{u}\in\R^{m\times d}$ and $\mac{w}\in\R^{n\times d}$ are orthonormal, and  $\mac{s}$ is a diagonal matrix. The truncated SVD gives
\begin{equation}
    \label{eq:trunsvdt}
    \mac{t} = \mac{m} \mac{w} = \mac{u} \mac{s}. 
\end{equation}
By Eckart-Young theorem \cite{36eckart}, matrix $\est{\mac{m}} = \mac{u}\mac{s}\mac{w}^T=\mac{t}\mac{w}^T$ is
the best rank $d$ approximation to matrix $\mac{m}$ with respect to the Frobenius
norm, i.e., the sum of squares of the entries of matrix $\mac{m} - \mac{x}$,
where $\mac{x}$ is of rank~$d$, attains its minimum at
$\mac{x}=\est{\mac{m}}$. Once we have approximated matrix \mac{m} with rank $d$
matrix $\mac{t}\mac{w}^T$, we can recall the procedure presented in the
previous subsection. Note that by having the SVD decomposition of matrix $\est{\mac{m}}$:
\begin{equation}
    \label{eq:estmsvd}
    \est{\mac{m}} = \underbrace{\mac{u}}_{\mac{e}}\underbrace{\mac{s}\mac{w}^T}_{\mac{b}},
\end{equation}
we can compute the analog of Eq.~\eqref{eq:bdef} by taking $\mac{e}=\mac{u}$ and $\mac{b}=\mac{s}\mac{w}^T$ as depicted in Eq.~\eqref{eq:estmsvd}.

The following algorithm concludes this subsection.
\begin{alg}\label{alg:noise}
    \par {\bf{Inputs:}} Sensor readings in the form of matrix $\mac{m}\in\R^{m\times n}$, the length $c$ of vectors $\kol[j]{v}$ (see Eqs.~\eqref{eq:readingmodel}--\eqref{eq:macn})
    {\bf{Output:}} Matrices $\est{\mac{v}}\in\R^{d\times n}$ and
    $\est{\mac{a}}\in\R^{d\times m}$ that, by the product
    $\est{\mac{a}}^T\est{\mac{v}}$, form the best rank $d$ approximation to
    matrix \mac{m}.
    \begin{enumerate}
        \item Compute truncated SVD of rank $d$ for matrix \mac{m}: 
            \[\mac{m}\approx  \mac{u}\mac{s}\mac{w}^T\]
        \item \label{step:two} Solve the following set of linear equations for the entries of
            symmetric matrix $\mac{g}\in\R^{d\times d}$:
            \[c^2 = \kol[j]{b}^T \mac{g} \kol[j]{b}, \quad j=1,\,\dots,\,n,\]
            where \kol[j]{b} are the columns of matrix $\mac{b}=\mac{s}\mac{w}^T$,
        \item Compute the eigendecomposition of matrix $\mac{g}$: 
            \[
                \mac{g} = \mac{q}^T \mac{\lambda} \mac{q}.
            \]
        \item Compute $\est{\mac{v}} = \mac{\lambda}^{\frac{1}{2}} \mac{q} \mac{b}$
            and $\est{\mac{a}} = \mac{\lambda}^{-\frac{1}{2}}\mac{q}\mac{u}^T$.
    \end{enumerate}
\end{alg}
Note that Algorithm~\ref{alg:noise} generalizes Algorithm~\ref{alg:noiseless},
i.e., Algorithm~\ref{alg:noise} may also be used in the absence of noise.
Also, Remarks~\ref{rem:ortho} and~\ref{rem:c} stay valid except for the
necessary change of the corresponding equalities that hold up to an orthogonal transformation into approximate equalities $\est{\mac{a}} \approx \mac{a}$ and $\est{\mac{v}}\approx \mac{v}$.
\begin{remark}
    If the sensor readings are noisy and the system of equations in
    Step~\ref{step:two} of Algorithm~\ref{alg:noise} is overdetermined, then the solution
    referred to in that step is meant to be the least square solution.
\end{remark}
\begin{remark}\label{rem:capprox}
    Algorithm~\ref{alg:noise} is robust to small divergence from the assumption
    on equal length of vectors \kol[j]{v}, as the disparity between the lengths
    of the vectors can be considered as an extra noise that contributes to
    $\mac{\eta}$ in Eq.~\eqref{eq:macn}.
\end{remark}

\section{The number of required measurements}\label{sec:positions}
As stated in Remark~\ref{rem:ortho}, Algorithm~\ref{alg:noiseless} reconstructs
matrices \mac{a} and \mac{v} up to an orthogonal transformation, provided that
Eq.~\eqref{eq:vnormf} has a unique solution~$\mac{g}$. The same holds for an approximate
reconstruction in the case of Algorithm~\ref{alg:noise}. Since matrix
$\mac{g}\in\R^{d\times d}$ is symmetric, the number of linear equations
required to specify the entries of \mac{g} uniquely is $\frac{d(d+1)}{2}$. In other words, in order to reconstruct matrices \mac{a} and \mac{v} with Algorithms~\ref{alg:noiseless} or~\ref{alg:noise} the number of positions $n$ has to satisfy the following inequality
\begin{equation}
    \label{eq:nreq}
    n \geq \frac{d(d+1)}{2}.
\end{equation}
In this section, we show that this bound cannot be loosened in general, i.e., there exist cases in which no algorithm can reconstruct matrices \mac{a} and \mac{v} with a smaller number of measurement setups~$n$.

The number of scalar measurements that form matrix \mac{p} is $nm$. The given magnitude of vectors $\wek{v}_i$ results in $n$ extra scalar data.
The number of unknown entries of matrices \mac{a} and \mac{v} is $d(n+m)$. These matrices are to be determined up to an orthogonal transformation of $\R^d$.
The dimension of the group of such transformations of $\R^d$ is $\frac{d(d-1)}{2}$. Thus,
for the desired reconstruction of \mac{a} and \mac{v}, the following inequality must hold
$
    nm +n \geq (m + n)d - \frac{d(d-1)}{2}.
    $
By rearranging this inequality, we get
\begin{equation}
    \label{eq:dimeq}
    \left( n - d  \right) \left( m - d + 1 \right) \geq \frac{d(d-1)}{2}.
\end{equation}
In particular, the number of setups $n$ hast to be greater than the dimension $d$, and the
number of sensors $m$ must be at least $d$. Moreover, for the minimal number of
sensors $m=d$, Inequality~\eqref{eq:dimeq} results in~\eqref{eq:nreq}.
In particular, $n \geq 3$ for dimension $d=2$, and 
$n \geq 6$ for $d=3$.
\begin{figure}[tpb]
    \centering
    \includegraphics[]{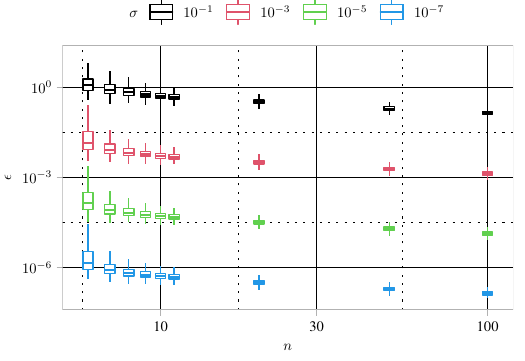}
    \caption{Calibration error of four triads of gyroscopes}
    \label{fig:boxplot1a}
\end{figure}
\section{Calibration of multiple gyroscopes}\label{sec:model1}
\begin{figure}[tpb]
    \centering
    \includegraphics[]{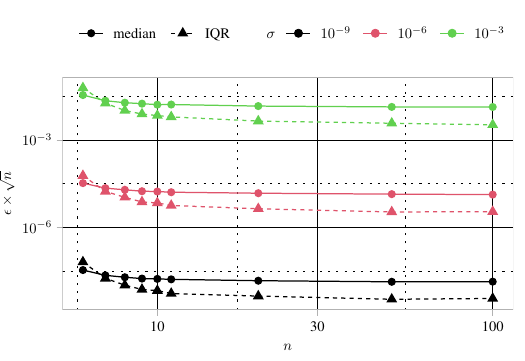}
    \caption{Calibration error statistics (for four triads of gyroscopes) as functions of the number of calibration positions $n$}
    \label{fig:line1}
\end{figure}

Gyroscopes are often produced in presumably orthonormal triads. 
Multiple instances of such triads can be used to reduce measurement errors
after data fusion~\cite{Wang21}.
We have considered a system of four gyroscope triads in the configuration shown in
Fig.~\ref{fig:model1}, i.e., with sensitive vectors of the gyroscopes constituting the following matrix:
\setcounter{MaxMatrixCols}{12}
\begin{equation*}
    \label{eq:gyrmaca}
    \mac{a}_{\text{model}} = 
    \begin{pmatrix}
        1 & 0 & 0 & 0 & 1 & 0 & 0 &-1 & 0 & 1 & 0 & 0\\
        0 & 1 & 0 &-1 & 0 & 0 & 1 & 0 & 0 & 0 & 0 &-1\\
        0 & 0 & 1 & 0 & 0 & 1 & 0 & 0 & 1 & 0 & 1 & 0
    \end{pmatrix}
.\end{equation*}
Such a system needs calibration because of the internal
sensitive-axes misalignment, scale factor spread between sensors comprising
every single triad, and the finite precision of the multi-triad
assembly. One may perform the needed calibration with the help of a rotary
table that can turn with a known angular rate. We propose the following
measurements for each of $n$ different positions of the system on the table.
\begin{enumerate}
    \item Place the system in the $i$-th position on the steady rotary table
        and record the reading of the sensors. The time-averaged readings $\wek{b}_i'$ are used as
        the estimates for the biases of the gyroscopes.
    \item Switch on the rotary table and wait until it rotates steadily.
    \item Record and time-average the readings of the sensors to form vectors $\wek{m}_i'$.
\end{enumerate}
We remove the measurement bias by subtracting the steady-state read-outs
from the readings obtained during the rotary movement, i.e., we set
\begin{equation}
    \label{eq:biasrem}
    \wek{m}_i = \wek{m}_i' - \wek{b}_i'.
\end{equation}
\begin{figure}[tpb]
    \centering
    \includegraphics[]{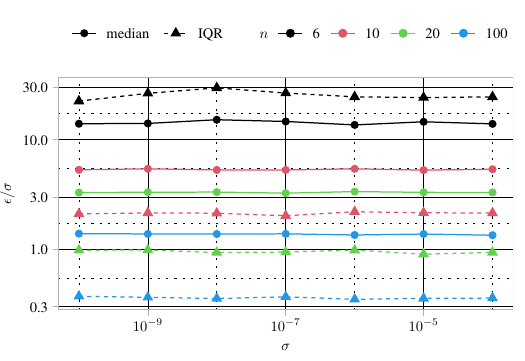}
    \caption{Calibration error statistics (for four triads of gyroscopes) as functions of the standard deviation~$\sigma$ of the measurement noise}
    \label{fig:line2a}
\end{figure}
To simulate a real scenario, we assumed that each of the sensor sensitivity
axes, represented by columns of matrix $\mac{a}$, differ from the columns of
$\mac{a}_{\text{model}}$ by a~random Gaussian vector with zero mean and diagonal covariance matrix with $\sigma=0.01$ on the diagonal. We picked
at random $n$ different positions (orientations) of the system and simulated the
sensors readings according to Eq.~\eqref{eq:macn}. Then, we invoked
Algorithm~\ref{alg:noise} to compute the matrix $\est{\mac{a}}$ of sensitive
vectors of the gyroscopes. Matrix $\est{\mac{a}}$ is expected to
approximate \mac{a} up to an orthogonal transformation. Therefore, to
compare these matrices, we first find two orthogonal matrices $\mac{h}$ and
$\mac{k}$ such that 
$\mac{h}\mac{a}$ and $\mac{k}\est{\mac{a}}$ are 
upper-triangular with positive entries on the main diagonal.
Then, we compute the calibration error $\epsilon$, which we define as the Frobenius norm of matrix
$\mac{h}\mac{a}-\mac{k}\est{\mac{a}}$, i.e.,
\begin{equation}
    \label{eq:frobnorm}
    \epsilon = \|\mac{h}\mac{a} - \mac{k}\est{\mac{a}}\|_F,
\end{equation}
where $\|\mac{x}\|_F$ denotes the square root of the sum of squares of the entries of matrix~\mac{x}.
For each of the considered values of $n$ and noise standard deviation $\sigma$, we
have repeated the simulation of the calibration procedure $1000$ times to assess
its statistical behavior. Fig.~\ref{fig:boxplot1a} shows the boxplot of the
calibration error. 
The line plots of Fig.~\ref{fig:line2a} indicate that
the median and the interquartile range of the calibration error are
approximately proportional to the standard deviation $\sigma$ of the measurement error. 
Figure~\ref{fig:line1} shows that these statistics drop with the number $n$ of
measurements at the rate $n^{-0.5}$ approximately.

\section{Conclusion}\label{sec:conclusion}
We have proposed a novel method for calibrating multiple-sensor systems with the
help of a constant magnitude vector quantity. The crucial merit of the method is the closed form for the computed parameters under calibration.
The method requires the sensors to be unbiased or the sensor bias to be computed before applying the proposed calibration procedure. We have noted that the
proposed algorithm allows for the calibration of the sensors up to an
orthogonal transformation, provided the magnitude of the measured quantity is known.
In the opposite case, the calibration procedure leaves a common scale factor
unresolved. We have deduced the minimum number of positions needed to complete
the calibration with the above-specified degree of ambiguity. The proposed calibration
method accuracy depends on the number of considered positions and the
measurement noise level. The results of the conducted numerical experiments
indicate that the calibration error is linearly proportional to the standard
deviation of measurement errors and inversely proportional to the square of the
number of positions. 

\bibliographystyle{IEEEtran}
\bibliography{IEEEabrv,biblio}

\end{document}